\documentclass[10pt]{iopart}
\usepackage{epsfig}
\usepackage{color}
\usepackage{colortbl}
\usepackage[table]{xcolor}
\usepackage{pxfonts}
\usepackage{mathrsfs}

\usepackage{amssymb}
\usepackage{graphicx}

\renewcommand{\rho}{\varrho}

\renewcommand{\e}{\mathtt{e}}
\newcommand{\p}{\wp}
\newcommand{\R}{\mathbf{R}}
\newcommand{\N}{\mathbf{N}}
\newcommand{\CC}{\mathscr{C}}
\newcommand{\pemp}{\wp_{\mathrm{emp}}}
\newcommand{\Pemp}{P_{\mathrm{emp}}}

\newcommand{\K}{\mathcal{K}}
\renewcommand{\d}{\mathtt{d}}
\newcommand{\dx}{\mathtt{d}x}

\newcommand{\dz}{\mathtt{d}z}

\newcommand{\dt}{\mathtt{d}t}
\newcommand{\dy}{\mathtt{d}y}

\newcommand{\tauin}{\tau^{(\mathrm{in})}}
\newcommand{\tauout}{\tau^{(\mathrm{out})}}

\newcommand{\argmin}{\mathrm{argmin}}
\newcommand{\argmax}{\mathrm{argmax}}
\newcommand{\supp}{{\mathrm{supp}}}

\newcommand{\TT}{\mathtt{T}}
\newcommand{\Flag}{\mathtt{F}}
\newcommand{\inb}{\mathrm{inb}}
\renewcommand{\kappa}{\varkappa}
\newcommand{\B}{{\mathscr B}}
\renewcommand{\S}{{\mathscr S}}
\newcommand{\E}{{\mathtt E}}
\renewcommand{\P}{{\mathbb P}}
\newcommand{\nbar}{{\overline{n}}}
\renewcommand{\geq}{\geqslant}
\renewcommand{\leq}{\leqslant}

\newcommand{\BE}{\begin{equation}}
\newcommand{\EE}{\end{equation}}

\begin{document}

\title[Vehicular headways on signalized intersections: theory, models, and reality]{Vehicular headways on signalized intersections: theory, models, and reality}

\author{Milan Krb\'alek and Ji\v r\'i \v Sleis}

\address{Faculty of Nuclear Sciences and Physical Engineering, Czech Technical University in Prague,
Prague, Czech Republic}
\ead{milan.krbalek@fjfi.cvut.cz}

\begin{abstract}
This article mediates an mathematical insight to the theory of vehicular headways measured on signalized crossroads.
Considering both, mathematical and empirical substances of the socio-physical system studied, we firstly formulate several theoretical and empirically-inspired criteria for acceptability of theoretical headway-distributions. Sequentially, the multifarious families of statistical distributions (commonly used to fit real-road headway statistics) are confronted with these criteria, and with original experimental time-clearances gauged among neighboring vehicles leaving signal-controlled crossroads after a green signal appears. Another goal of this paper is, however, to decide (by means of three completely different numerical schemes) on the origin of statistical distributions recorded by stop-line-detectors. Specifically, we intend to examine whether an arrangement of vehicles is a consequence of traffic rules, driver's estimation-processes, and decision-making procedures or, on contrary, if it is a consequence of general stochastic nature of queueing systems. \end{abstract}


\maketitle

\section{Introduction}

Due to the practical background, modeling of spatial positions of vehicles in the vicinity of signalized intersection attracts a permanent interest of researches. As is intelligible, the detection of statistical distributions of spatial/time headways among vehicles may lead to a more accurate determination of intersection capacities, which finally results in an economic profit. Indeed, the importance of the topic is noticeable from a high frequency of recent scientific papers dealing with intersection analyses. However, a majority of those works focus on macro-description (for example, \cite{Chung,Chopard,Brockfeld,Fouladvand-01}) or (if concentrating on micro-structure) on average values of traffic micro-quantities (for example, \cite{Abdelwahab,Fouladvand-02,Fouladvand-03,America}). In the last years there have been published many studies investigating some detailed statistical distributions of inter-vehicular headways (spatial or time) between neighboring cars occurring close the stop line. Some of them focus on distribution of departure time intervals (\cite{Jin_and_Zhang,Wang,Li_and_Wang}), other on distribution of spatial gaps between cars waiting for a green signal (\cite{Red_cars,Chen-Li-Zhang}).

\begin{figure}[htb]
\begin{center}
\epsfig{file=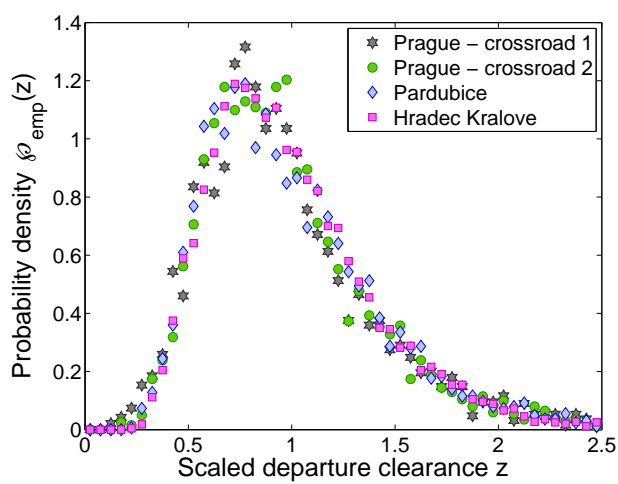,height=2.5in}
\begin{flushright}\parbox{15.0cm}{\caption{The empirical histogram of departure clearances. The constituent signs represent the statistical frequency of scaled netto-time gaps among neighboring cars leaving the intersections located as announced in the legend. \label{fig:empirical_histograms}}}
\end{flushright}
\end{center}
\end{figure}

In this article we intend to analyze larger amount of original individual data gauged on various crossroads (located in Czech cities Praha, Pardubice, and Hradec Kr\'alov\'e) and to introduce suitable theoretical predictions for relevant probability density of vehicular micro-quantities. Moreover, our aim is to create numerical representations of crossroad models leading to statistically consistent distributions. Finally, the nominated analytical clearance-distributions will be confronted with theoretical criteria derived from a short-ranged trait of mutual vehicular interactions. In the last part of the text we will try to insight into the nature of the examined distributions.

\section{Empirical departure-clearance statistics}

The vehicular data analyzed in this work were gauged on multi-lane intersections located in the centers of Czech cities Praha (Prague), Pardubice and Hradec Kr\'alov\'e. All the tested intersections are constituents of an extensive network of roads and crossroads inside the internal metropolis and are therefore strongly saturated. In all cases the time interval between two green signals (on one crossroad) is short, which means that some cars are not able to reach the threshold of the following intersection (during one green phase) and therefore have to wait for another green light. This fact finally leads to a substantial decrease in average speed of cars moving among crossroads, i.\,e. one can observe here the effects detectable ordinarily in congested traffic regimes (see \cite{Review-Helbing,Review-Kerner}).

\begin{figure}[htb]
\begin{center}
\epsfig{file=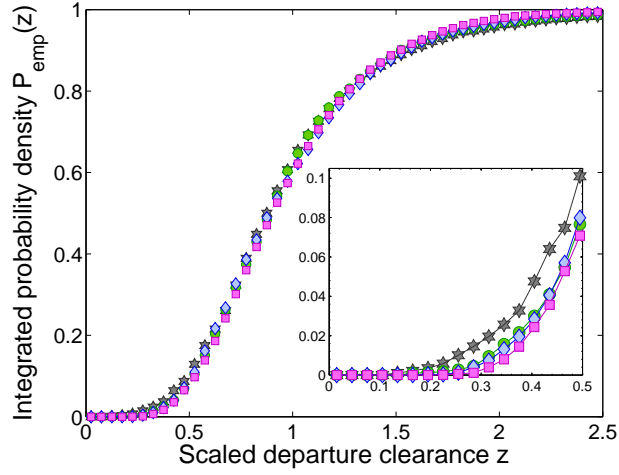,height=2.5in}
\begin{flushright}\parbox{15.0cm}{\caption{The cumulated histogram of departure time-clearances. The constituent signs represent the cumulated probability density for scaled netto-time gaps detected among neighboring cars leaving the intersections located as announced in figure \ref{fig:empirical_histograms}. The behavior of the cumulative distribution function near the origin is magnified in the inset. \label{fig:empirical_histograms_cumulated}}}
\end{flushright}
\end{center}
\end{figure}

The traffic experiment has been organized as follows. The spontaneous traffic flow near the chosen intersection (see table \ref{Tab:zkusebni}) has been controlled by traffic lights in a usual mode. No external interventions has been applied. The gauging procedure (i.\,e. measurement of departure times $\tauin_k$ and $\tauout_k$ -- see the mathematical notation below) had started in the moment of replacing the red signal by the green one and finished immediately after another red signal. We add that all analyzed traffic quantities have been measured only at such intersections where other cars (moving in different lanes or in different arms) do not influence the gauged cars.

Thus, let the symbols $\tauin_k$ and $\tauout_k$ indicate the times when the front/back bumpers of $k$th car $(k=0,1,2,\ldots,N)$  have intersected a reference line (stop line, typically) at the chosen intersection threshold. Then the \emph{time clearance} between succeeding cars is defined as \BE t_k:=\tauin_k-\tauout_{k-1}\quad (k=1,2,\ldots,N). \EE
The fundamental quantity analyzed in our article is $z_k=t_k/\bar{t},$ where $\bar{t}=\sum_{k=1}^N t_k/N,$ and referred to as the \emph{scaled time clearance.} The empirical probability density $\pemp(z),$ being usually plotted as staircase function, is then called the \emph{(scaled) time clearance distribution}. To eliminate an unwelcome dependency of empirical distributions on the binning (quantization of detected data into given smaller intervals -- bins) one can define the integrated probability density (cumulative distribution function)
\BE\Pemp(z)=\int_{-\infty}^z \pemp(y)\,\dy.\EE
The general quantitative results of a preliminary statistical analysis of gauged traffic data are summarized in table \ref{Tab:zkusebni}, where it is visible that the mean clearance is about 1.6 seconds (with a standard deviation equal approximately to 0.7\,$sec$). Here we remark that the quantity measured for purposes of this article (clearance -- netto gap) is different from the quantity (brutto headway) analyzed in the research paper \cite{Jin_and_Zhang}.

\begin{table}
\caption{\label{Tab:zkusebni} Evaluation of data records before the scaling procedure.}
\begin{indented}
\lineup
\item[]\begin{tabular}{clccc}
\br
\rowcolor[gray]{.9}  Number & Location & Sample size & Mean clearance & Variance\\
\mr
1 & Prague - crossroad 1 & 3785  & 1.6237\,$sec$ & 0.6649\,$sec^2$\\
2 & Prague - crossroad 2 & 4022  & 1.5226\,$sec$ & 0.5023\,$sec^2$\\
3 & Pardubice            & 3279  & 1.6110\,$sec$ & 0.4997\,$sec^2$\\
4 & Hradec Kr\'alov\'e   & 8795  & 1.5820\,$sec$ & 0.4185\,$sec^2$\\
\br
\end{tabular}
\end{indented}
\end{table}

\section{Criteria for acceptability of analytical clearance-distributions}\label{SEC:Criteria for acceptability}

Owing to an empirical background of the topic investigated in this research the curves representing theoretical approximations of the intersection-clearance-distribution have to fulfill both, the mathematical and empirically-inspired criteria. Whereas mathematical criteria are deduced from exact theoretical definitions, empirical criteria reflects realistic features of traffic microstructure distributions. The measure for acceptability of theoretical curves can be therefore quantified by a number of fulfilled criteria.

First of all, we briefly summarize mathematical criteria. If $\p(z)$ is intended to be declared a theoretical prediction for time-clearance distribution, it should meet the following theoretical criteria:
\BE \mathbf{(T1) - positivity:} \quad \forall z\in\R:~ \p(z)\geq 0,\label{T1-positivity}\EE
\BE \mathbf{(T2) - support~constraint:} \quad \supp(\p)=(0,\infty),\label{T2-positivity-of-support}\EE
\BE \mathbf{(T3) - normalization:} \quad \int_{\R} \p(z)\,\dz=1,\label{T3-normalization}\EE
\BE \mathbf{(T4) - scaling:} \quad \int_{\R} z\, \p(z)\,\dz=1\label{T4-scaling},\EE
\BE \mathbf{(T5) - continuity:} \quad \p(z)\in\CC(\R^+).\label{T5-continuity}\EE
We remark that the scaling criterion \textbf{T4} can be understood as optional.

\begin{figure}[htb]
\begin{center}
\epsfig{file=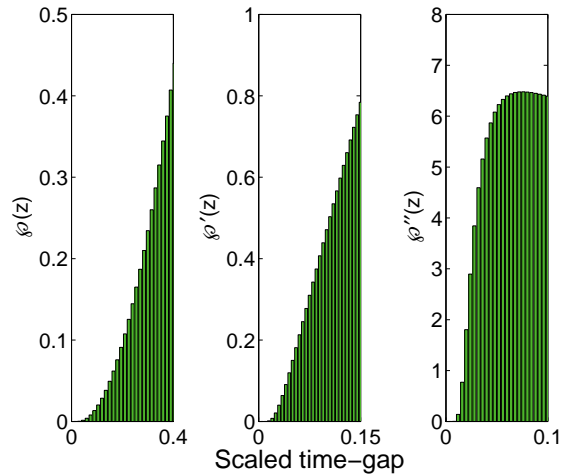,height=2.5in}
\begin{flushright}\parbox{15.0cm}{\caption{The graphical visualization of the origin plateau in the empirical clearance distributions. The bars display the smoothed probability density for short traffic clearances, and the first and second derivatives. The analyzed data (for traffic densities between 40 and 60 vehicles per kilometer) have been extracted from extensive data samples gauged on the two-lane freeway D1 (the Czech Republic). \label{fig:derivative_near_the_origin}}}
\end{flushright}
\end{center}
\end{figure}

Except these properties some other requirements can be derived from recent knowledge about microscopic structure of vehicular samples. As is apparent from many scientific sources (see \cite{Cecile,Wang,Buckley,XiLiRiuXin,Kerner-Klenov-Hiller,KRB-Kybernetika,Review-Helbing,Tilch,Traffic_NV,Jin_and_Zhang}) the spatial or temporal headway/clearance-distributions (analyzed for congested traffic streams) show the heavy plateau located near the origin (see the figure \ref{fig:derivative_near_the_origin} and the inset of the figure \ref{fig:empirical_histograms_cumulated}). Such a plateau can be understood as a consequence of strong repulsions among closely-occurring vehicles whose drivers make an effort to prevent a possible crash. Mathematically, such a phenomenon is described by means of the following definition:
\BE \mathbf{(E1) - the~origin~plateau:} \quad \forall m\in\N:~ \lim_{z\rightarrow 0_+} z^{-m}\p(z)=0,\label{E1-platea}\EE
which is (for the locally smooth densities $\p(z)\in\CC^\infty(0,\delta)$) equivalent to the conditions $\frac{\d^m \p}{\d z^m}(0_+)=0.$ Unfortunately, $\p(z)$ is not (as immediately follows from the preceding) an analytical function, which therefore does not allow its Taylor's expansion.

The second empirically-inspired criterion is induced from the perspicuous fact that all vehicular interactions are short/middle-ranged, i.\,e. movements of two sufficiently outlying cars are not correlated (even in the congested traffic). Such statistical ensembles used to be usually referred to as \emph{quasi-poissonian.} This terminology reflects the general knowledge that a system is qualified as  \emph{poissonian (purely poissonian)} if all associated subsystems are independent. In this case, the probability for occurrence of several elements inside the fixed (space or time) region conforms to a Poissonian distribution. If the interaction among elements are restricted to several neighbors only, the poissonian features of adjacent elements are broken. On contrary, outlying elements behave still as independent, which leads to the similarity between the distribution tail and that derived for poissonian ensembles. Therefore, the tails of related headway distributions (for pure and quasi poissonian ensembles, respectively) show the similar trends. These considerations result in the undermentioned definition.

\textbf{Definition \ref{Criteria for acceptability}.1} A probability density $\p(z)$ (and the associated distribution function) is called \emph{balanced} if there exists  $\omega>0$ so that
\BE \forall \kappa>\omega:\quad \lim_{z\rightarrow +\infty} \p(z)\e^{\kappa z}=+\infty, \EE
and
\BE \forall \kappa\in(0,\omega):\quad \lim_{z\rightarrow +\infty} \p(z)\e^{\kappa z}=0. \EE
The number $\omega$ is then called \emph{the balancing index} and denoted by $\inb(\p).$ The class of balanced distributions is denoted by $\B.$

As it is evident, the family of balanced distributions and the family of heavy-tailed distributions (see \cite{Cline,Teugels}) are disjoint. Thus, the intersection of $\B$ and the class $\S$ of subexponential distributions is empty. The same holds true also for class of fat-tailed or long-tailed distributions. Therefore the class $\B$ is a special subclass of light-tailed distributions. Based on an assumption that vehicular interactions are short/middle-ranged, the empirical netto-time gap distributions should also meet (see \cite{My_Multiheadways} as well) the final criterion:
\BE \mathbf{(E2) - the~balanced~tail:} \quad \p(z)\in\B.\label{E2-balanced-tail}\EE

\section{Functional candidates for time clearances distributions}\label{Criteria for acceptability}

\begin{figure}[htb]
\begin{center}
\epsfig{file=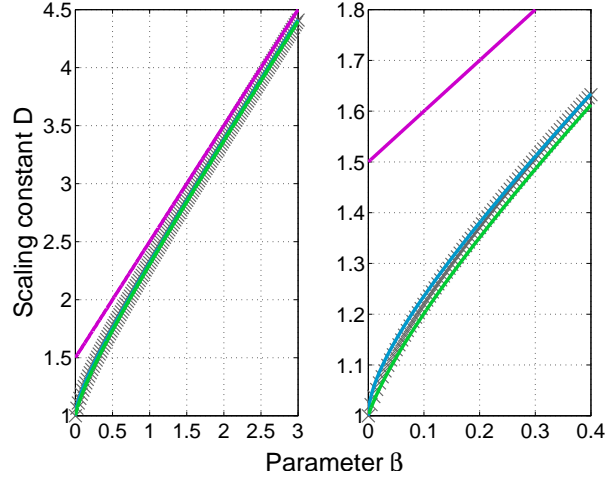,height=2.5in}
\begin{flushright}\parbox{15.0cm}{\caption{The calibration of the scaling constant $D=D(\beta)$ in the Generalized inverse Gaussian distribution. The red, blue, and green curves represents the asymptotical dependency (\ref{dance_with_me}), phenomenological approximation  (\ref{dance_with_you}), and analytical approximation (\ref{dance_with_nobody}), respectively. The crosses display numerical solutions of the scaling equality $\int_\R \wp_{\mathtt{GIG}}(z)\,\dz=\int_\R z\wp_{\mathtt{GIG}}(z)\,\dz.$ \label{fig:normalizacni-konstanta}}}
\end{flushright}
\end{center}
\end{figure}

With the respect to the previous explorations of empirical clearances near signalized intersections \cite{Li_and_Wang,Wang,Red_cars,Jin_and_Zhang,Tolle,Luttinen} the following non-composite distribution models can be applied for describing the real-road headway statistics: \emph{Exponential Distribution, Erlang Distribution, Nakagami Distribution \cite{Nakagami}, Log-Normal Distribution, and Generalized inverse Gaussian distribution \cite{GIG}.} We emphasize that the exponential, Erlang, and Generalized inverse Gaussian distributions represent (contrary to the Nakagami and Log-Normal Distribution) a theoretically-reasoned probability densities. Indeed, their forms have been derived as steady-state distributions for a local thermodynamic ensemble with short-ranged repulsions among the elements (see \cite{Krbalek_gas,Helbing_and_Krbalek}). After the scaling procedure, all these distributions read as
\BE \wp_{\mathtt{EXP}}(z)=\Theta(z)\e^{-z}, \label{Exponential Distribution} \EE
\BE \wp_{\mathtt{ERL}}(z)=\Theta(z)\frac{(\omega+1)^{\omega+1}}{\Gamma(\omega+1)}z^\omega\e^{-(\omega+1)z}, \label{Erlang Distribution} \EE
\BE \wp_{\mathtt{NAK}}(z)=2\Theta(z)z^{2m-1}\frac{\Gamma^{2m} \left(m+\frac{1}{2}\right)}{\Gamma^{2m+1}(m)}\exp\left[-\frac{\Gamma^{2} \left(m+\frac{1}{2}\right)}{\Gamma^{2}(m)}z^2\right], \label{NakagamiDistribution} \EE
\BE \wp_{\mathtt{LN}}(z)=\frac{\Theta(z)}{\sqrt{2\pi}\sigma z}\exp\left[-\frac{(\sigma^2+2\ln(z))^2}{8\sigma^2}\right]. \label{Log-Normal Distribution} \EE
Similarly, also the additional probability density $\wp_{\mathtt{GIG}}(z)=A\exp[-\beta/z-Dz]$ (considered in the articles \cite{Krbalek_gas,Red_cars,KRB-Kybernetika,Traffic_NV,Review-Helbing} and analyzed in the book \cite{GIG}) requires the proper normalization and scaling. Owing to the functional relation \BE \int_0^\infty \e^{-\frac{x^2}{4t}}\e^{-t}\,\dt=x\K_1(x),\EE where $\K_1(x)$ stands for the Macdonald's function of the first order -- solution of the modified Bessel's differential equation of the second kind (of the order $\alpha\in\N$) $x^2y''+xy'-(x^2+\alpha^2)y=0,$ one can derive the value of the normalization constant
\BE A^{-1}= 2\sqrt{\frac{\beta}{D}}\K_1(2\sqrt{\beta D}).\EE
Substituting $z(x):=x^\alpha \e^x \K_\alpha(x)$ into the original Bessel's equation we obtain the differential equation $xz''-(2\alpha +2x-1)z'+(2\alpha-1)z=0$ which (together with the Cauchy's initial conditions $z(0)=z'(0)=(2\alpha-2)!!$)  provides a more suitable small-$x$ approximation
\BE \K_\alpha(x) \approx (2\alpha-2)!!\left(1+\frac{2x}{2\alpha-1}\right)^{\alpha-1/2}\frac{\e^{-x}}{x^\alpha} \label{aproximace_nova} \EE
than the well-known approximation $\K_\alpha(x) \approx \e^{-x}/x^\alpha .$ Since the above-mentioned normalization integrals (\ref{T3-normalization}) and (\ref{T4-scaling}) fulfill the differential equation
\BE \int_0^\infty x\e^{-\frac{\nu^2}{x}}\e^{-\varkappa^2 x}\,\dx=-\frac{1}{2\varkappa} \frac{\partial}{\partial \varkappa} \int_0^\infty \e^{-\frac{\nu^2}{x}}\e^{-\varkappa^2 x}\,\dx, \EE
the scaling condition $\mathbf{(T4)}$ can be reformulated (applying the approximation (\ref{aproximace_nova})) into the cubic equation
\BE 4\nu\varkappa^3+(1-4\nu^2)\varkappa^2-4\nu\varkappa-1=0.\EE
Its real solution then provides a desired functional relation guaranteeing a fulfilment of the scaling condition. Such a relation is of a form
\BE D \approx \left(\frac{4\beta+w(\beta)+\frac{16\beta^2+40\beta+1}{w(\beta)}-1}{12\sqrt{\beta}}\right)^2, \label{dance_with_nobody} \EE
where $w^3(\beta)=4\left(16 \beta^3+60 \beta^2+3   \sqrt{48 \beta^3+132\beta^2-3\beta}+39\beta \right)-1.$ Asymptotical features of the normalization dependency $D=D(\beta)$ may be quantified by the relations
\BE \lim_{\beta \rightarrow 0_+} D(\beta)=1, \quad D(\beta) \approx \beta+\frac{3}{2}~ (\beta \gg 1). \label{dance_with_me}\EE
Accuracy of the previous approximate calculations is demonstrated in figure \ref{fig:normalizacni-konstanta} where the numerically-specified values $D$ are confronted to the analytically- and phenomenologically-specified values. To conclude, one can briefly summarize that the relation
\BE D(\beta)\approx \beta+\frac{3-\e^{-\sqrt{\beta}}}{2} \label{dance_with_you} \EE
represents a sufficient approximation of the scaling constant in Generalized inverse Gaussian distribution, which means that the probability density
\BE \wp_{\mathtt{GIG}}(z)=\frac{\sqrt{D}\,\Theta(z)}{2\sqrt{\beta}\K_1(2\sqrt{\beta D})}\exp\left[-\frac{\beta}{z}-Dz\right], \quad D=\beta+\frac{3-\e^{-\sqrt{\beta}}}{2} \label{GIG Distribution} \EE
completes the set of non-composite probabilistic models convenient for purposes of this work.

\begin{table}
\caption{\label{Tab:srovnani-vlastnosti} Criteria of acceptability for various non-composite probabilistic models.}
\begin{indented}
\lineup
\item[]\begin{tabular}{cccccccc}
\br
\rowcolor[gray]{.9}  Probability &  Criterion &  Criterion &  Criterion &  Criterion &  Criterion & Criterion & Criterion \\
\rowcolor[gray]{.9} Density &  $\mathbf{(T1)}$ &  $\mathbf{(T2)}$ &  $\mathbf{(T3)}$ &  $\mathbf{(T4)}$ &  $\mathbf{(T5)}$ & $\mathbf{(E1)}$ & $\mathbf{(E2)}$ \\
\mr
$\wp_{\mathtt{EXP}}(z)$ & yes & yes & yes & yes & yes & no & yes \\
$\wp_{\mathtt{ERL}}(z)$ & yes & yes & yes & yes & yes & no & yes \\
$\wp_{\mathtt{NAK}}(z)$ & yes & yes & yes & yes & yes & no & no \\
$\wp_{\mathtt{LN}}(z)$  & yes & yes & yes & yes & yes & yes & no \\
$\wp_{\mathtt{GIG}}(z)$ & yes & yes & yes & yes & yes & yes & yes \\
\br
\end{tabular}
\end{indented}
\end{table}

In table \ref{Tab:srovnani-vlastnosti} we summarize the relevant properties of all above-mentioned distributions. As is apparent, the one and only probabilistic model fulfilling all the requisite criteria is the model derived as a steady-state solution for the thermal-like vehicular simulator presented in the articles \cite{Krbalek_gas} and \cite{Helbing_and_Krbalek}. Other distributions show at least one incompatibility with theoretical requirements. However, all suggested functions can be used for comparing with empirical clearance distributions gauged between neighboring vehicles leaving a chosen signal-controlled intersection. For these purposes we define the generalized statistical distance
\BE \chi(\varepsilon)=\int_0^\infty \bigl|\wp(z;\varepsilon)-q(z)\bigr|^2 z\e^{1-z}\,\dz \label{statistical-distance}\EE
cumulating the weighted deviations between a theoretical one-parametric prediction $\wp(z;\varepsilon)$ and empirical frequency $q(z).$ The optimal value of the estimated parameter $\hat{\varepsilon}$ can be then enumerated by minimizing the statistical distance (\ref{statistical-distance}), i.\,e.
\BE \hat{\varepsilon} = \argmin_{\varepsilon\in[0,\infty)} \int_0^\infty \bigl|\wp(z;\varepsilon)-q(z)\bigr|^2 z\e^{1-z}\,\dz. \label{Miluna} \EE
The tangible results of such procedure are tabularized in tables \ref{Tab:fitovane-parametry} and \ref{Tab:statisticke-vzdalenosti} where optimal values of the estimated parameters are summarized as well as minimal values of weighted statistical distances (\ref{statistical-distance}) specified for the above-mentioned optimal parameters. We remark that the weight factor $\phi(z)=z\exp[1-z]$ has been chosen \emph{(a)} to eliminate an influence of long clearances, \emph{(b)} to suppress extremely short clearances, and finally \emph{(c)} to increase an influence of clearances being close to the mean value. In addition, $\argmax_{z\geq 0} \phi(z)=1$ and $\phi(1)=1.$

\begin{table}
\caption{\label{Tab:fitovane-parametry} Optimal values of parameters for various one-parametric probabilistic models.}
\begin{indented}
\lineup
\item[]\begin{tabular}{ccccccc}
\br
\rowcolor[gray]{.9}  Location & $\wp_{\mathtt{ERL}}(z)$ & $\wp_{\mathtt{NAK}}(z)$ & $\wp_{\mathtt{LN}}(z)$ & $\wp_{\mathtt{GIG}}(z)$\\
\mr
Prague - crossroad 1 & $\hat{\omega}=4.8350$ & $\hat{m}=1.6619$ & $\hat{\sigma}=0.41931$ & $\hat{\beta}=2.0507$\\
Prague - crossroad 2 & $\hat{\omega}=5.3017$ & $\hat{m}=1.7939$ & $\hat{\sigma}=0.40551$ & $\hat{\beta}=2.2489$\\
Pardubice            & $\hat{\omega}=4.6100$ & $\hat{m}=1.6116$ & $\hat{\sigma}=0.43015$ & $\hat{\beta}=1.9172$\\
Hradec Kr\'alov\'e   & $\hat{\omega}=5.4536$ & $\hat{m}=1.8281$ & $\hat{\sigma}=0.39985$ & $\hat{\beta}=2.3195$\\
\br
\end{tabular}
\end{indented}
\end{table}

\begin{table}
\caption{\label{Tab:statisticke-vzdalenosti} Statistical distances (\ref{statistical-distance}) for various one-parametric probabilistic models.}
\begin{indented}
\lineup
\item[]\begin{tabular}{cccccccc}
\br
\rowcolor[gray]{.9}  Location & $\chi_{\mathtt{EXP}}(z)$  & $\chi_{\mathtt{ERL}}(z)$ & $\chi_{\mathtt{NAK}}(z)$ & $\chi_{\mathtt{LN}}(z)$ & $\chi_{\mathtt{GIG}}(z)$\\
\mr
Prague - crossroad 1 & $5.9368$ & $0.43766$ & $0.83113$ & $0.24556$ & $0.27659$\\
Prague - crossroad 2 & $6.5770$ & $0.35958$ & $0.75371$ & $0.14868$ & $0.17275$\\
Pardubice            & $5.9281$ & $0.28204$ & $0.61882$ & $0.11643$ & $0.11985$\\
Hradec Kr\'alov\'e   & $6.5628$ & $0.13128$ & $0.42344$ & $0.02901$ & $0.02859$\\
\br
\end{tabular}
\end{indented}
\end{table}

\section{Rigidity of quasi-poissonian ensembles}\label{sec:Rigidity}

\begin{figure}[htb]
\begin{center}
\epsfig{file=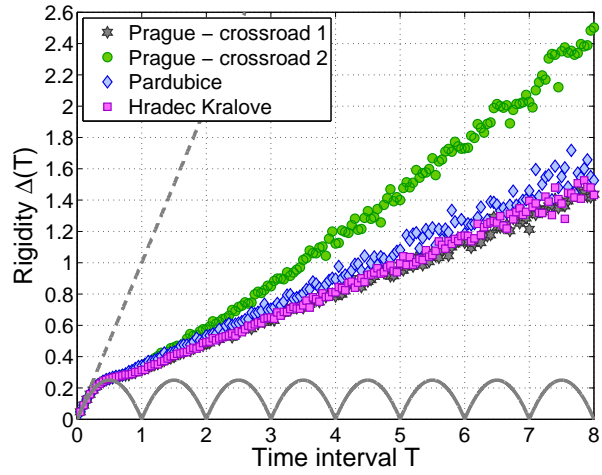,height=2.5in}
\begin{flushright}\parbox{15.0cm}{\caption{Rigidity of empirical traffic data. The signs represent statistical rigidity $\Delta(T)$ analyzed for clusters of cars leaving the intersections located according to the legend. The dashed line and wavy line symbolize the statistical rigidities calculated for ensembles of uncorrelated (\ref{uncorrel-rigidity}) or equidistantly spaced (\ref{equi-rigidity}) particles, respectively.  \label{fig:rigiditka}}}
\end{flushright}
\end{center}
\end{figure}

Configuration of vehicles in a intersection neighborhood used to be typically analyzed, as discussed in the previous sections, by statistical instruments applied to gaps or time-intervals between departures of succeeding cars (see \cite{Jin_and_Zhang,Wang,Li_and_Wang,Red_cars,Chen-Li-Zhang}). Although recent researches have proposed certain candidates for distance/time clearance distributions, the way how to evaluate such probabilistic models is still missing. Concurrently, the felicitous evaluation-scheme can be found in Random Matrix Theory \cite{Mehta}. Here a mathematical quantity (called \emph{spectral rigidity} or \emph{number variance}) is defined. This quantity surveys a structure of eigenvalue-clusters in ensembles of random matrices. The noticeable advantages of such an approach are as follows: 1. The spectral rigidity quantifies (contrary to the clearance distribution) an arrangement of larger clusters of particles/cars/eigenvalues. 2. The functional formula for the rigidity is directly connected to the clearance distribution, which could bring an interesting alternative how to verify newly-suggested probabilistic predictions against empirical data. 3. Geometric shapes of rigidity curves are extremely simple. 4. Statistical analysis of spectral rigidity for data files is undemanding. 5. A slight change in the clearance distribution leads to a marked change in the graph of rigidity, which demonstrates a strong sensitivity of rigidity-testing. This effect is noticeably visible in figures \ref{fig:empirical_histograms_cumulated} and \ref{fig:rigiditka} (compare circles and diamonds in both figures).

If reformulated within the bounds of traffic theory the rigidity coincides with the following interpretation. Consider a set $\{z_i:i=1 \ldots N\}$ of scaled time-clearances between each pair of subsequent cars. Since we suppose that the mean time gap taken over the complete set is re-scaled to one, it holds $\sum_{i=1}^N z_i=N.$ After dividing the time interval $[0,N)$ into subintervals $[(k-1)T,kT)$ of a length $T$ one can define a new random variable $n_k(T)$ representing the number of cars whose departure-times belong to the $k-$th subinterval. The average value $\nbar(T)$ taken over all possible subintervals is therefore
\BE \nbar(T)=\frac{1}{\lfloor N/T \rfloor} \sum_{k=1}^{\lfloor N/T \rfloor} n_k(T)=T, \EE
where the integer part $\lfloor N/T \rfloor$ stands for the number of all subintervals $[(k-1)T,kT)$ included in the entire interval $[0,N).$ We suppose, for convenience, that $N/T$ is integer. The \emph{time rigidity} $\Delta(T)$ is then defined as
\BE \Delta(T)=\frac{T}{N} \sum_{k=1}^{N/T} \bigl(n_k(T)-T\bigr)^2. \label{definice-rigidity} \EE
Providing that all variables are independent (which is not the general case) the formula (\ref{definice-rigidity}) represents the statistical variance of the number of vehicles passing a given fixed point (a threshold of the intersection, typically) during the time interval $T.$ With the respect to the fact, that expected value $\E(n(T))$ and average value $\nbar(T)$ can differ (for systems of depending random variables) we will not use the term "variance." It is self-evident that for ensembles of equidistantly spaced elements the statistical rigidity reads
\BE \Delta_{\mathtt{EQD}}(T)=\bigl(T-\lfloor T \rfloor\bigr)\bigl(\lfloor T \rfloor+1-T\bigr). \label{equi-rigidity} \EE

Denoting $\wp_\ell(z)$ distribution of netto-time gaps between $\ell+2$ cars (i.\,e. $\wp_0(z)=\wp(z)$ is the standard clearance distribution) one can define the \emph{cluster function}
\BE R(z)=\sum_{\ell=0}^\infty \wp_\ell(z) \label{cluster-function} \EE
that is closely related to the random variable $n(T)$ -- number of particles departing a chosen location during the time interval $T.$  Indeed, probability $\P[n(T)=\ell],$ that exactly $\ell$ cars pass the stop-line during arbitrary time-interval of length $T,$ can be expressed in terms of  multi-clearance distributions $\wp_\ell(z)$ as
\begin{eqnarray}
\P\bigl[n(T)=0\bigr]=1-\int_0^T \wp_0(z)\,\dz,\nonumber \\ \P\bigl[n(T)=\ell\bigr]=\int_0^T \bigl(\wp_{\ell-1}(z)-\wp_\ell(z)\bigr)\,\dz.
\end{eqnarray}
Hence the average value of $n(T)$ is
\BE \E\bigl(n(T)\bigr)=\sum_{\ell=0}^\infty \ell\, \P[n(T)=\ell]=\int_0^T R(z)\,\dz. \EE
Furthermore,
\BE \E\left(n^2(T)\right)=\sum_{\ell=0}^\infty \ell^2 \P[n(T)=\ell]=\int_0^T \left(2S(z)-R(z)\right)\,\dz, \EE
where $S(z)= \sum_{\ell=0}^\infty \ell \wp_\ell(z).$ Since $R(z) \star R(z)=S(z)-R(z)$ (as follows from rules derived for functional convolutions) the rigidity can be computed via
\begin{eqnarray} \Delta(T)=\sum_{\ell=0}^\infty (\ell-T)^2 \P[n(T)=\ell] \nonumber \\=2\int_0^T (R \star R)(z)\,\dz+(1-2T)\int_0^T R(z)\,\dz+T^2.\end{eqnarray}
Assuming an approximate equality $\E(n(T))\approx\nbar(T),$ i.\,e. $\int_0^T R(z)\,\dz\approx T,$ and using the convolution property $\int_0^T (R \star R)(z)\,\dz=R(T) \star\int_0^T R(z)\,\dz$ we obtain the closing formula for the statistical rigidity
\BE   \Delta(T) \approx T-2\int_0^T (T-z)(1-R(z))\,\dz. \label{rigidity-final}\EE

According to articles \cite{Bogomolny,Traffic_NV} the Laplace image of cluster functions of quasi-poissonian ensembles analyzed in our paper, i.\,e. ensembles with the balanced tails (see the definition 4.1), read
\BE \mathscr{L}[R_{\mathtt{EXP}}](p)=\frac{1}{p}, \EE
\begin{eqnarray}
 \mathscr{L}[R_{\mathtt{ERL}}](p)=\frac{1}{\left(1+\frac{p}{\omega+1}\right)^{\omega+1}-1} \nonumber \\ 
 =\frac{1}{p}-\frac{\omega }{2 (\omega +1)}+\frac{p \omega\left(\omega+2 \right)}{12 (\omega +1)^2}+\mathcal{O}(p^2),
\end{eqnarray}
\begin{eqnarray} 
\mathscr{L}[R_{\mathtt{GIG}}](p) \approx \left(\frac{D+p}{D}~\frac{\e^{2\sqrt{(D+p)\beta}}}{\e^{2\sqrt{D\beta}}}-1\right)^{-1} \nonumber\\ \approx \frac{1}{p}-\frac{2D\beta+3\sqrt{D\beta}}{4\bigl(1+\sqrt{D\beta}\bigr)^2}+ p\frac{6\sqrt{D\beta}+D\beta\bigl(21+4D\beta+16\sqrt{D\beta}\bigr)}{48D\bigl(1+2\sqrt{D\beta}\bigr)^3}+\mathcal{O}(p^2).
\end{eqnarray}
On the account that the formula (\ref{rigidity-final}) can be rewritten to the form
$$\Delta(T) \approx  T - 2\Theta(T)T \star \Theta(T) + 2\Theta(T)T \star R(T),$$
the linear trend $\Delta(T)\approx \lambda T + \mu$ near infinity may be revealed (after applying the Laplace transformation) with the help of
$$\lambda p + \mu p^2 \approx p-2+2p\mathscr{L}[R](p).$$
Taylor's expansion to the function $p\mathscr{L}[R](p)$ then finalizes the process of rigidity-linearization. Whence the linear tails of the adjoint rigidities are given by
\begin{eqnarray}
\Delta_{\mathtt{EXP}}(T)=T,\label{uncorrel-rigidity}\\
 \Delta_{\mathtt{ERL}}(T) \approx \frac{T}{\omega +1}+\frac{\omega\left(\omega+2 \right)}{6(\omega +1)^2},\\
\Delta_{\mathtt{GIG}}(T) \approx \frac{2+\sqrt{D\beta}}{2D\bigl(1+\sqrt{D\beta}\bigr)}T + \frac{6\sqrt{D\beta}+D\beta\bigl(21+4D\beta+16\sqrt{D\beta}\bigr)}{24\bigl(1+2\sqrt{D\beta}\bigr)^4}.
\end{eqnarray}

As is well known from the Random Matrix Theory, the linear asymptote $\Delta(T)\approx \lambda T + \mu$ (characterizing the course of rigidity near infinity) demonstrates the short-ranged nature of component interactions, which is in a consonance with the general meaning on driver's interactions. Really, if investigating the statistical rigidity in vehicular samples one can detect typical linear tails in all examined data-samples (see figure \ref{fig:rigiditka}). Moreover, as expected, the rigidities of all presented probabilistic models show the linear tails, and for that reason one can compare the related slopes $\lambda$ with those obtained by analysing empirical data. The quantitative outcome of such a comparison is synoptically summarized in table \ref{Tab:sloupy-rigidit} where we show the values of the rigidity-slopes $\lambda$ obtained for the parameters (and models)  summarized in table \ref{Tab:fitovane-parametry}. For completeness, we add that the statistical rigidities for exponential, Erlang, and GIG distributions has been acquired analytically, whereas Nakagami and Log-Normal rigidities has been observed numerically.

\begin{table}
\caption{\label{Tab:sloupy-rigidit} The slope $\lambda$ in the theoretical and empirical rigidities.}
\begin{indented}
\lineup
\item[]\begin{tabular}{ccccccccc}
\br
\rowcolor[gray]{.9}  Location & $\lambda_{intersection}$ & $\lambda_{\mathtt{EXP}}$  & $\lambda_{\mathtt{ERL}}$ & $\lambda_{\mathtt{NAK}}$ & $\lambda_{\mathtt{LN}}$ & $\lambda_{\mathtt{GIG}}$\\
\mr
Prague - crossroad 1 & $0.1625$ & $1.0000$ & $0.1821$ & $0.1706$ & $0.1930$ & $0.1873$\\
Prague - crossroad 2 & $0.3048$ & $1.0000$ & $0.1617$ & $0.1474$ & $0.1732$ & $0.1738$\\
Pardubice            & $0.1815$ & $1.0000$ & $0.1816$ & $0.1720$ & $0.2074$ & $0.1961$\\
Hradec Kr\'alov\'e   & $0.1691$ & $1.0000$ & $0.1573$ & $0.1516$ & $0.1658$ & $0.1692$\\
\br
\end{tabular}
\end{indented}
\end{table}

\section{Assessment of suggested probabilistic models}\label{sec:Evaluation}

In the previous sections we have suggested and evaluated five probability distributions that are broadly accepted as analytical candidates for vehicular headway distributions. As demonstrated by the previous quantitative and qualitative evaluations, the choice of candidates is effortlessly tenable. However, the selected evaluation criteria give stronger preference to the Log-Normal and Generalized inverse Gaussian models, because both of them fit the empirical histograms so that the statistical distance (\ref{statistical-distance}) is rapidly smaller than that enumerated for Exponential, Erlang, and Nakagami models. Furthermore, also the associated rigidities (empirical and Log-Normal/GIG) are in a plausible correspondence. Taking into consideration the theoretical and empirically-inspired criteria ($\mathbf{T1-T5}$ and $\mathbf{E1-E2}$) we can convincingly conclude that the best theoretical predictions for vehicular departure-times has been achieved by means of probability density (\ref{GIG Distribution}). In the last resort, we remind that the great advantage of GIG-model is also the fact that the proposed density is of a socio-physical essence. Indeed, the distribution (\ref{GIG Distribution}) has been identified in the articles \cite{Krbalek_gas,Helbing_and_Krbalek,KRB-Kybernetika,My_Multiheadways} as a steady-state distribution of a certain socio-physical traffic model. In addition to that, the book \cite{GIG} points out that probability density (\ref{GIG Distribution}) characterizes a distribution of times between events in some renewal processes. These findings support the final result of our evaluation-procedure.

\section{Leave-the-intersection models: GCF scheme}\label{sec:GCF-Model}

In the following three sections we will propose three traffic models aiming to explain the core of departure clearance distributions, especially to reproduce the observed vehicular gap distributions on signal-controlled crossroads. First of them is based on the car-following principles discussed in \cite{Gazis,Brackstone,Li_and_Wang,Jin_and_Zhang,XiLiRiuXin} and on the theory of the so-called Galton's board (see \cite{Galton}). Such a model will be referred to as \emph{GCF model.} Parameters of the model and their brief explanations are summarized in table \ref{Tab:parametrouskove}.

\begin{table}
\caption{\label{Tab:parametrouskove} Parameters of \emph{GCF model.}}
\begin{indented}
\lineup
\item[]\begin{tabular}{cccc}
\br
\rowcolor[gray]{.9}  Nomenclature & General Extent & Option & Description\\
\mr
$w_{\mathtt{start}}$ & $\in [2,3]~ m/s$ & $2.7~ m/s$ & velocity delimiting the starting mode\\
$w_{\mathtt{max}}$ & $\in [10,20]~ m/s$ & $16~ m/s$ & maximal velocity\\
$a_{\mathtt{start}}$ & $\in [2,5]~ m/s^2$ & $2~ m/s^2$ & run-up acceleration\\
$a_{\mathtt{plus}}$ & $>a_{\mathtt{start}}$ & $4.4~ m/s^2$ & maximal free-driving acceleration \\
$a_{\mathtt{minus}}$ & $\approx 2a_{\mathtt{plus}}$ & $7~ m/s^2$ & maximal braking deceleration \\
$g_{\mathtt{start}}$ & $\in [1,3]~ m$ & $2.2~ m$ & minimal distance required for moving off\\
$g_{\mathtt{min}}$ & $\in (0,1]~ m$ & $0.5~ m$ & minimal safety clearance\\
$g_{\mathtt{max}}$ & $>10~ m$ & $15~ m$ & distance limit for free-driving mode\\
$H$ & $\in [1,10]~ s$ & $8~s$ & deceleration time\\
$p$ & $\in (0,1)$ & $0.38$ & random-deceleration rate \\
$\vartheta$ & $\in (0,1)$  & $0.8$ & decelerating factor \\
\br
\end{tabular}
\end{indented}
\end{table}

Consider $N$ identical dimensionless particles located at the time $t=0$ in a sequentially organized locations $x_N<x_{N-1}<\ldots<x_2<x_1<0.$ Here the origin $x=0$ represents a intersection threshold.  For brevity of following notations, we denote the space headway in front of $\ell$th particle as $r_\ell,$ i.e., $r_\ell:=x_{\ell-1}-x_{\ell}. $ At the beginning of each realization of the GCF algorithm, the initial velocities $v_1,v_2,\ldots,v_N \geq 0$ of all particles are set for zero and initial positions are randomized according to selected distribution. Copying the approach in \cite{Jin_and_Zhang} we introduce a Boolean variable $\Flag_\ell$ signalizing if the $\ell$th vehicle is in the starting-up mode ($\Flag_\ell=1$) or not.

The simulating scheme is divided into three main modes: \emph{(1) stopped mode, (2) starting-up mode,} and \emph{(3) moving mode}. The latter is composed from three sub-model:  \emph{(3a) free-driving sub-mode, (3b) braking sub-mode,} and \emph{(3c) car-following sub-mode}. Dynamical rules for transition of ensemble from an original state (at the time $t$) to an updated state (at the time $t+\TT,$ where $\TT$ denotes the simulation time-span) are then strictly derived from the above-mentioned modes.

\subsection{Stopped mode}

Entering condition for this mode is $v_\ell(t)=0.$ The $\ell$th vehicle will enter the starting-up mode at the time $t+\TT$ (and $\Flag_\ell(t+\TT)$ will be set for one) if the $r_\ell(t)>g_{\mathtt{start}}$ otherwise the vehicle will continue in the actual mode.

\subsection{Starting-up mode}

If $\Flag_\ell(t)=1$ the move of the $\ell$th vehicle will be regulated by the starting-up rule. If $v_\ell(t) \geq w_{\mathtt{start}}$ then the $\ell$th vehicle will enter the moving mode at the time $t+\TT$ and $\Flag_\ell(t+\TT)$ will be set for zero. On contrary, if $v_\ell(t)<w_{\mathtt{start}}$ then $\Flag_\ell(t+\TT):=0$ and
\BE v_\ell(t+\TT):= \Theta(r_\ell(t) - g_{\mathtt{min}})(v_\ell(t)+a_{\mathtt{start}}\TT), \EE
where $\Theta(x)$ is the Heaviside's step-function.

\subsection{Free-driving sub-mode}

If and only if $v_\ell(t) \geq w_{\mathtt{start}}$ and the distance headway  $r_\ell(t)$ becomes larger than the distance limit $g_{\mathtt{max}},$ the vehicle enter the free-driving sub-mode. Then
\BE v_\ell(t+\TT):= \min\{v_\ell(t)+a_{\mathtt{plus}}\TT,w_{\mathtt{max}}\}. \EE

\subsection{Braking sub-mode}

If and only if $v_\ell(t) \geq w_{\mathtt{start}},$ $r_\ell(t)<g_{\mathtt{max}},$ and \BE v_\ell(t)-v_{\ell-1}(t)> \frac{r_\ell(t) - g_{\mathtt{min}}}{H},\EE
there exists a risk of collision. Therefore the velocity must be reduced as
\BE v_\ell(t+\TT):= \max\{v_\ell(t)-a_{\mathtt{minus}}\TT,0\}. \EE

\begin{figure}[htb]
\begin{center}
\epsfig{file=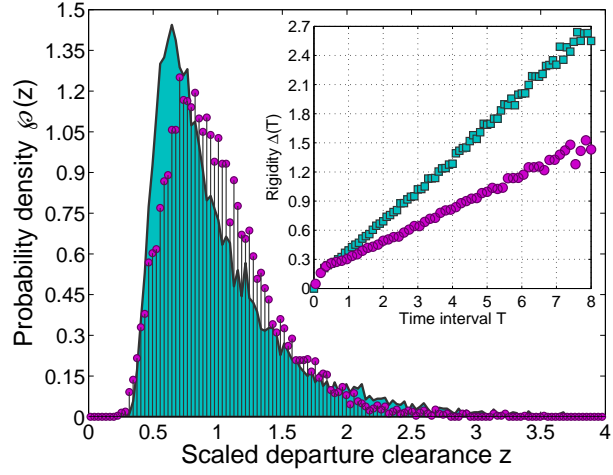,height=2.5in}
\begin{flushright}\parbox{15.0cm}{\caption{Clearance distributions and statistical rigidities for the GCF model. The main plot compares the clearance distributions between real-road data (Hradec Kr\'alov\'e -- circles) and Galton-inspired car-following model (area-plot) presented in the text. The comparison between statistical rigidities (for the same data ensembles) is presented in the inset.  \label{fig:modely-cinske}}}
\end{flushright}
\end{center}
\end{figure}

\subsection{Car-following sub-mode}

If and only if $v_\ell(t) \geq w_{\mathtt{start}},$ $r_\ell(t)<g_{\mathtt{max}},$ and \BE v_\ell(t)-v_{\ell-1}(t) \leq \frac{r_\ell(t) - g_{\mathtt{min}}}{H},\EE the driver carefully adapts his/her maneuvering to a previous car. Specifically, the Galton-inspired stochastic update-rule for the car-following process (see \cite{Galton,Li-Tau}) is introduced:
\BE v_\ell\left(t+\frac{\TT}{2}\right) := \left\{\begin{array}{cccc} \frac{r_\ell(t)}{r_\ell(t-\TT)}\vartheta v_\ell(t) & \ldots & \mathrm{with~probability} & p, \\ \max\left\{\frac{r_\ell(t)}{r_\ell(t-\TT)}\frac{v_\ell(t)}{\vartheta},w_{\mathtt{start}}\right\} & \ldots &  \mathrm{with~probability} & 1-p,\end{array}\right. \EE
\BE v_\ell\left(t+\TT\right) := \min \left\{ v_\ell(t)+a_{\mathtt{plus}}\TT; v_\ell\left(t+\frac{\TT}{2}\right) \right\}\EE

\subsection{Forward-ordered update}

Finally, the positions of particles are sequentially updated (in forwardly directed order) as
\BE x_\ell(t+\TT)= x_\ell(t)+ \TT \cdot v_\ell(t+\TT). \EE

The above-mentioned update rules, defining the forward-ordered sequential dynamics of the system, have been repeatedly applied to actual configurations until the last car has intersected the stop line. Denoting the time, when the $\ell$th car has reached the measuring point $x=0,$ as $\tau_\ell,$ one can intuitively define the scaled clearances as
\BE z_\ell=(N-1)\frac{\tau_\ell-\tau_{\ell-1}}{\tau_N-\tau_1} \quad (\ell\in \{2,3,\ldots, N\}). \label{numericke-hedveje} \EE
These normalized time-gaps are independent of the time-span $\TT$ and represent the main quantities investigated in this paper. Therefore, after repeated realizations of the GCF algorithm (with the initialization: $N=12$ and $\TT=0.3$) we can proceed to an expected statistical evaluation. The graphical outputs of such evaluation are visualized in figure \ref{fig:modely-cinske}. Here one can compare the clearance distributions between empirical and GCF data (10\,000 clearances), as well as the respective statistical rigidities. Although certain similarities can be found there, deviations between model and traffic reality are relatively significant.

\section{Leave-the-intersection models: PLCF scheme}\label{sec:MCF-Model}

The second design for an intersection model (here called \emph{phenomenological car-following model}) is based on the work \cite{Krauss}. Similarly to the previous section, we firstly introduce all parameter of the model suggested (see table \ref{Tab:parametrouskove_02}).

\begin{figure}[htb]
\begin{center}
\epsfig{file=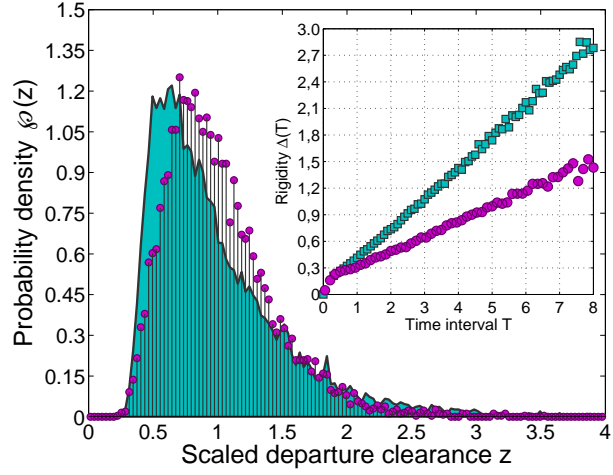,height=2.5in}
\begin{flushright}\parbox{15.0cm}{\caption{Clearance distributions and statistical rigidities for the PLCF model. The main plot compares the clearance distributions between real-road data (Hradec Kr\'alov\'e -- circles) and phenomenological car-following model (area-plot) presented in the text. The comparison between statistical rigidities (for the same data ensembles) is presented in the inset.  \label{fig:modely-nemecke}}}
\end{flushright}
\end{center}
\end{figure}

Our PLCF-modification eliminates a slight illogicality in the original conception presented in \cite{Krauss}. Namely, an occurrence of two vehicles moving with the same velocities and with zero clearance (i.e., cars moving like connected objects) is in fact extremely improbable. Therefore, we eliminate such a circumstance by introducing the minimal (i.e., safety) bumper-to-bumper distance $g_{\mathtt{min}}$ that is randomly chosen from the exponential distribution $\mathtt{Exp}(\varepsilon)$ with parameter $\varepsilon>0.$ Now, the simulation scheme replicates the general strategy of the GCF model. Specifically, we consider $N$ particles placed in locations $x_N<x_{N-1}<\ldots<x_2<x_1<0$ and moving with velocities $v_1,v_2,\ldots,v_N \geq 0.$ Again, bumper-to-bumper distances are denoted by $r_\ell.$ Furthermore, we define the so-called \emph{safe velocity}
\BE v_{\mathtt{safe}_\ell}(t) = v_{\ell-1}(t) + \frac{r_\ell(t)-g_{\mathtt{min}_\ell}-\TT \cdot v_{\ell-1}(t)}{\frac{v_\ell(t)+v_{\ell-1}(t)}{2 a_{\mathtt{minus}}}+\TT },\EE
whose rigorous form is derived (see \cite{Krauss}) by requiring a collision-free condition and limitedness of vehicular accelerations. By means of the safe-velocity-approach we can express the \emph{desired velocity} as
\BE v_{\mathtt{des}_\ell}(t)=\min\{w_{\mathtt{max}};v_{\mathtt{safe}_\ell}(t); v_\ell(t)+\TT \cdot a_{\mathtt{plus}}\}.\EE
Then the randomly perturbed velocity (influenced by the phenomenological coefficient $\theta$ suppressing the velocity-variance in the ensemble) satisfies the equation
\BE v_\ell(t+\TT)= \max\left\{0;\mathtt{Uni}\bigl(v_{\mathtt{des}_\ell}(t)-\theta\cdot\TT \cdot a_{\mathtt{minus}}),v_{\mathtt{des}_\ell}(t)\bigr)\right\}, \EE
where the symbol $\mathtt{Uni}(a,b)$ corresponds to the continuous uniform distribution on the interval $(a,b).$ Finally, the positions of particles are standardly updated (in forwardly directed order, again) according to
\BE x_\ell(t+\TT)= x_\ell(t)+ \TT \cdot v_\ell(t+\TT). \EE
The outputs of the PLCF model (obtained for the fixed initialization conditions: $N=12$ and $\TT=0.2$ and for a calibrated value of the suppress coefficient $\theta$) are then subjected to standard statistical tests analyzing a microstructure of the particle ensemble. The results of those tests are plotted in figure \ref{fig:modely-nemecke}.

\begin{table}
\caption{\label{Tab:parametrouskove_02} Parameters of \emph{PLCF model.}}
\begin{indented}
\lineup
\item[]\begin{tabular}{cccc}
\br
\rowcolor[gray]{.9}  Nomenclature & General Extent & Option & Description\\
\mr
$w_{\mathtt{max}}$ & $\in [10,20]~ m/s$ & $14~ m/s$ & maximal allowable velocity\\
$a_{\mathtt{plus}}$ & $\in [3,5]~ m/s^2$ & $4.35~ m/s^2$ & maximal free-driving acceleration \\
$a_{\mathtt{minus}}$ & $\in [4,8]~ m/s^2$ & $7~ m/s^2$ & maximal braking deceleration \\
$\theta$ & $\in [1/2,1]$ & $0.7$ & suppress coefficient \\
$g_{\mathtt{min}_\ell}$ & $>0~ m$ & $\in \mathtt{Exp}(2/3)$ &  individual safety clearance\\
\br
\end{tabular}
\end{indented}
\end{table}

\section{Leave-the-intersection models: annealing-based scheme (AB scheme)}\label{sec:Sleis-Model}

The intention of our article is, inter alia, to examine whether the arrangement of vehicles in the vicinity of an intersection is a consequence of traffic rules, complicated evaluation-procedures, and sophisticated decision-making procedures inside a driver's brain or, on contrary, it is a consequence of general stochastic nature of queueing systems. For solving this dilemma we intend to create a stochastic alternative for both above-discussed models. Thus, we will introduce an unimodal scheme simulating a time-evolution of vehicular ensembles without any division into modes (contrary to the GCF model) and without a concept of the safe values for some quantities (contrary to the PLCF model).

For these purposes we have created an original model based on principles of the so-called simulated annealing \cite{Scharf}. We consider $N$ dimensionless particles located along the ring with a circumference equal to $N.$ Initial locations of particles are generated equidistantly in the interval $[0,N-X],$ where $X$ represents a free gap before the leading vehicle (typically, the distance to the rear of a queue waiting on a previous intersection). Relative velocities $v_\ell(t=0)$ of all vehicles are reset. Then the repeating procedure is applied as follows.

\begin{enumerate}
\item Timing is shifted by one.
\item Quasi-energy of the ensemble is calculated via \BE E(t)=\sum_{\ell=1}^{N-1} \frac{1}{x_\ell(t)-x_{\ell+1}(t)}. \label{kvasimodo} \EE
\item An index $\ell\in\{1,2,\ldots,N\}$ is picked at random.
\item Relative velocity $v_\ell$ is updated: $v_\ell(t+1)=\min\{v_\ell(t)+1/m,1\},$ where $m\in\N$ is the fixed parameter (see table \ref{Tab:parametrouskove_03}).
\item Using formula \BE U_\ell= \frac{\eta}{x_\ell(t)-x_{\ell+1}(t)}  + \frac{1}{x_{\ell-1}(t)-x_\ell(t)}\EE the individual quasi-potential of the $\ell-$th vehicle is calculated. We remark that the coefficient $\eta$ reduces a influence of a vehicle behind.
\item A random number $\delta \backsim \mathtt{Uni}(0,1)$ is drawn and an anticipated position \BE x_\ell(t+1)=x_\ell(t)+\delta w_{\mathtt{max}} v_\ell(t+1) \EE of $\ell$th element is computed.
\item As the vehicles can not change their order we accept $x_\ell(t+1)$ only if $x_\ell(t+1)<x_{\ell-1}(t).$ Moreover, if $x_\ell(t+1) \geq x_{\ell-1}(t)$ then the relative velocity should be reduced according to $v_\ell(t+1):=\max\{0,v_\ell(t+1)-1/m.\}$
\item Potential \BE U^\prime_\ell= \frac{\eta}{x_\ell(t+1)-x_{\ell+1}(t)}  + \frac{1}{x_{\ell-1}(t)-x_\ell(t+1)} \EE of new configuration is enumerated.
\item If $U^\prime_\ell \leq U_\ell$ the $\ell$th particle position takes on a new value $x_\ell(t+1).$
\item If $U^\prime_\ell > U_\ell$ then the Boltzmann factor $\hslash=\exp\left[-\gamma \Delta U\right],$ where $\Delta U=U^\prime_\ell - U_\ell,$ should be compared with another random number $r \backsim \mathtt{Uni}(0,1).$ Provided that
the inequality $\hslash> r$ is fulfilled the $\ell$th particle position takes on the new value $x_\ell(t+1)$ too. Otherwise, the original configuration remains unchanged, i.e. $x_\ell(t+1)=x_\ell(t).$ In this case, the relative velocity is reduced again: $v_\ell(t+1):=\max\{0,v_\ell(t+1)-1/m.\}$
 \end{enumerate}

\begin{table}
\caption{\label{Tab:parametrouskove_03} Parameters of \emph{AB model.}}
\begin{indented}
\lineup
\item[]\begin{tabular}{cccc}
\br
\rowcolor[gray]{.9}  Nomenclature & General Extent & Option & Description\\
\mr
$N$ & $\in \N $ & $36$ & number of vehicles\\
$m$ & $\in \N$ & $10$ & number of divisions in velocity discretization \\
$w_{\mathtt{max}}$ & $\in [2/m,2]$ & $0.7$ & maximal allowable velocity\\
$\gamma$ & $\in [0,+\infty)$ & $8.15$ & randomization parameter \\
$\eta$ & $\in [0,1]$ & $0.3$ &  reduction coefficient (reduces an influence of rear gaps) \\

$X$ & $\in (0,N)$ & $10.8$ &  effective distance between intersections \\
$\xi$ & $(N-X)/(N-1)$ & $0.72$ &  average gap among neighbors in an initial state  \\
\br
\end{tabular}
\end{indented}
\end{table}

\begin{figure}[htb]
\begin{center}
\epsfig{file=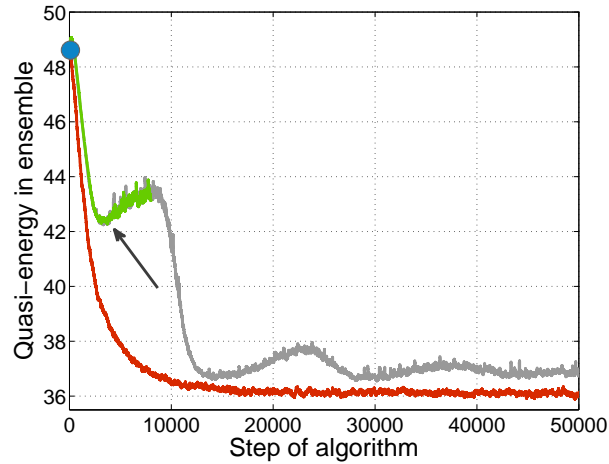,height=2.5in}
\begin{flushright}\parbox{15.0cm}{\caption{Time evolution of quasi-energy during the AB algorithm.
We plot the average value of quasi-energy (\ref{kvasimodo}) calculated for 500 repeated realizations of AB algorithm (green curve). For comparison purposes we also display (see the red curve) the evolution of energy in a classical variant of the annealing procedure (the so-called \emph{Metropolis algorithm}) simulating a transition of thermal gases into the thermodynamical equilibrium.
Gray curve demonstrates how the quasi-energy (\ref{kvasimodo}) develops if applied more updates than 8\,000. Blue circle represents the initial quasi-energy $E_{\mathtt{ini}}=(N-1)^2/(N-N\xi),$ whereas the gray arrow shows where the leading car has reached the last car waiting at a following intersection. \label{fig:kvazienergie}}}
\end{flushright}
\end{center}
\end{figure}

Although the classical scheme of the simulated annealing ensures a relaxation of ensemble into a thermal equilibrium (see \cite{Scharf}), here we are focused on non-equilibrium states of the above-mentioned  particle-ensemble. Furthermore, the introduced rules modify the original Metropolis algorithm so dramatically that even if the energy in a system had been established standardly, the proposed scheme would not lead to a state corresponding to a classical balance. Those facts are clearly understandable from the figure \ref{fig:kvazienergie}, where we investigate the time evolution of the energy (\ref{kvasimodo}) in the ensemble. Thus, after 8\,000 steps (when 15 cars have left the intersection) the system is still significantly far from any equilibrium, which is in full consonance with realistic situation of vehicles.

\begin{figure}[htb]
\begin{center}
\epsfig{file=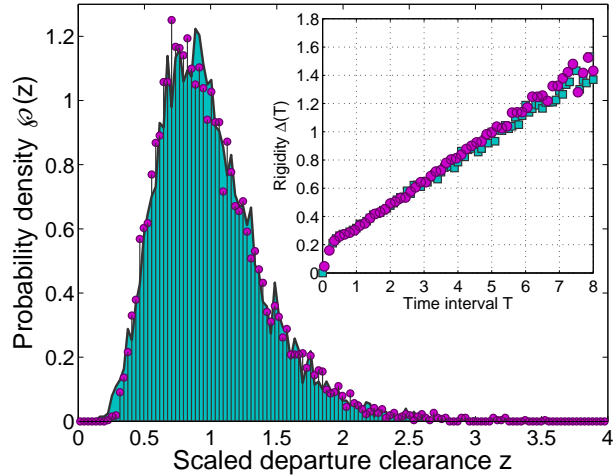,height=2.5in}
\begin{flushright}\parbox{15.0cm}{\caption{Clearance distribution and statistical rigidity for the AB model. The main plot compares the clearance distributions between Hradec Kr\'alov\'e data (red circles) and  the annealing-based model (area-plot). The comparison between statistical rigidities (for the same data ensembles) is visualized in the inset.  \label{fig:modely-sleisovske}}}
\end{flushright}
\end{center}
\end{figure}

Five hundreds repetitions of that scheme have then generated sufficient amount of inter-vehicle intervals suitable for intended statistical evaluations. Non-equilibrium distributions of re-scaled time-gaps in the suggested model (visualized in figure \ref{fig:modely-sleisovske}) illustratively demonstrate a more significant compliance with real-road statistics than those detected in the previous two models. Similarly, also the test of the statistical rigidity (lucidly shown in the inset of the figure \ref{fig:modely-sleisovske}) confirmed that the similarity between the AB model and intersection-reality is not accidental.

\section{Discussion and conclusion remarks}

This paper deals with theoretical and empirical background of vehicular dynamics investigated in the vicinity of signal-controlled intersections. Such a specific area of traffic research exploits a simplicity of inter-vehicle interactions near the traffic lights to a deeper understanding of general laws in vehicular dynamics. Indeed, some complicated traffic phenomena are there suppressed, which brings an unique opportunity for a disclosure of the nature of the issue examined.

Since some features of driving behavior are easily predictable (e.g. middle-ranged nature of mutual interactions) one can formulate certain theoretically-substantiated properties of statistical distributions for microscopic vehicular quantities. Using also the well-known empirical regularities in microscopic structure of traffic samples we have therefore formulated several criteria for acceptability of mathematical curves proposed for fitting empirical histograms. Sequentially, these criteria can serve to measure a quality of suggested statistical models. Such evaluations have been tested on several families of distributions in the fourth section. As is evident from these tests, some previously-proposed functions are not suitable as headway-statistics estimators. On contrary, Generalized inverse Gaussian distribution (\ref{GIG Distribution}) (fulfilling all the acceptance criteria) represents a relevant theoretical prediction for empirical departure-headway-statistics. Moreover, these findings have been supported by theoretical and empirical study of the associated statistical rigidities.

Detailed dynamics of vehicles passing the stop-line at signalized intersection has been analyzed by means of three simulation schemes (based on three different approaches). Although all those microscopic simulators have produced similar departure statistics, the comprehensive analyses (tests of the statistical rigidity, especially) have uncovered some serious discrepancies. The ability to reproduce empirical features of time intervals between two subsequent cars has been confirmed only for the non-equilibrium model based on principles of the simulated annealing. In this case, the consistency between empirical and numerically-obtained headways has also been accompanied by a correspondence between both rigidities.

However, the final outcome of our considerations about the origin of empirical headway-distributions is, in fact, extremely surprising. According to our observations, the models with more conspicuous stochastic component (as AB-scheme) produce more relevant predictions than models accenting certain interaction rules and traffic modes (as GCF/PLCF-schemes). For this reason, it can be speculated that the stochastic component of the examined system dominates the interaction-rules as well as decision-making processes inside the driver's brain. Furthermore, it has been demonstrated that original arrangement of vehicles (before the green signal appears) is stochastically perturbed in an extremely short time. This fact is clearly visible in AB-simulator, where original equidistant-sequencing of vehicles (characterized by a wavy curve of the statistical rigidity) is very quickly transformed into the stochastic sequencing (characterized by a linear rigidity being significantly distant from the above-mentioned wavy curve).  Also the time-dependence of quasi-energy shows the sharpest changes immediately after the beginning of the simulation. All these facts assure us that the decisive factor for movement of vehicular ensembles (near the stop line) is its stochasticity.

To conclude, this paper together with the article \cite{Red_cars} mediate a relatively comprehensive view into the spatio-temporal course of vehicular ensembles leaving a signalized intersection.

\subsection*{Acknowledgments}

The authors would like to thank the students of Faculty of Nuclear Sciences and Physical Engineering, Czech Technical University in Prague (attending the seminar 01SMB${}_2$ on mathematical applications, academic year 2010/2011) and Mr. Pavel K\v riv\'a\v n who gauged the traffic data analyzed in this paper. Also our thanks go to Mr. V\'it Hanousek who designs an original computer tool suitable for all the above-discussed measurements.  This work was supported by the Czech Technical University within the project SGS12/197/OHK4/3T/14.

\end{document}